\documentclass[pra,showpacs,preprint]{revtex4}
\usepackage{bm}
\usepackage{amsmath}
\usepackage{amssymb}
\usepackage{epsfig}

\def\be{\begin{equation}}
\def\en#1{\label{#1}\end{equation}}
\def\vec#1{\overrightarrow{#1}}

\newcommand{\eqb}{\begin{eqnarray}}
\newcommand{\eqe}{\end{eqnarray}}

\newcommand{\bx}{{\bf x}}

\newcommand{\bD}{{\bf D}}
\newcommand{\bb}{{\bf b}}
\newcommand{\bQ}{{\bf Q}}
\newcommand{\bq}{{\bf q}}
\newcommand{\br}{{\bf r}}
\newcommand{\bp}{{\bf p}}
\newcommand{\bP}{{\bf P}}
\newcommand{\ze}{{\bf 0}}
\newcommand{\bd}{{\bf d}}

\begin{document}
\title{ State engineering of Bose-Einstein condensate   in the  optical lattice by a  periodic sublattice of dissipative sites }
\author{ V. S. Shchesnovich  }
\affiliation{  Centro de Ci\^encias Naturais e Humanas, Universidade Federal do
ABC, Santo Andr\'e,  SP, 09210-170 Brazil}

\begin{abstract}
We introduce the  notion  of   dissipative periodic lattice as an optical lattice
with  periodically distributed  dissipative sites and argue that it allows to
engineer unconventional Bose-Einstein superfluids  with the complex-valued  order
parameter. We consider two examples,  the one-dimensional dissipative optical
lattice, where each third site is dissipative, and    the dissipative honeycomb
optical lattice, where each dissipative lattice site neighbors three non-dissipated
sites.  The tight-binding approximation is employed, which allows one to obtain
analytical results. In the one-dimensional case the condensate is driven to a
coherent Bloch-like state with non-zero quasimomentum, which breaks the
translational periodicity of the dissipative lattice. In the two-dimensional case
the condensate is driven to a zero quasimomentum Bloch-like state, which is a
coherent superposition of four-site discrete vortices of alternating vorticity with
the vortex centers  located at the dissipative sites.

\end{abstract}
 \maketitle
\section{Introduction}

Bose-Einstein condensates (BEC) in the optical lattices have  become   an universal
laboratory to study various quantum  phenomena of the condensed matter physics
\cite{PS,JZ,MO,Bloch}. A spectacular achievement in this direction is  the
demonstrated  single-site addressability of the two-dimensional optical lattices
\cite{ContRem1,ContRem2}, i.e.  the atoms can be  controllably removed from the
lattice sites selected at will.  Another intriguing recent development  is
observation of the unconventional  boson superfluidity in the higher bands of the
optical lattice \cite{Pbos,Fbos,SpinBos}, described by a complex-valued order
parameter and, therefore, beyond the  no-node theorem of R. Feynman (see for
instance, the review \cite{Wu}).

The purpose of the present paper is to demonstrate that the single-site
addressability in the optical lattices allows one to generate  the unconventional
boson superfluids. To this goal, one can use the controlled removal of atoms by the
technique of Refs.~\cite{ContRem1,ContRem2} from a periodic sublattice of the
optical lattice  with BEC loaded in it. It is shown below that  the condensate is
driven to a coherent non-local Bloch-like state described by a complex-valued order
parameter (in the two-dimensional case) or the order parameter with   nodes (in the
one-dimensional case). This is achieved by the emulation of a coherent non-local
dissipation by a local one, similar as in the  recently proposed scheme for
generators of  the non-classical states of photons \cite{MS,QSG}.

The quantum state engineering based on  the dynamics in the open quantum systems
\cite{QSEQC}, in particular in the optical lattices \cite{QSE}, is a well-known
idea. As distinct from the previous proposal of the quantum state engineering in
the optical lattices (see also the recent study \cite{DrDissLatt}), where the
inter-site atomic currents are coupled to reservoirs, in our case  the dissipation
acts \textit{locally and incoherently} on a periodic sublattice of the lattice
sites, whereas the condensate is driven to a non-local coherent state.

It is known that the action of   dissipation in conjunction with the nonlinearity
can increase   coherence of the quantum state of BEC. For instance, the possibility
to engineer   the order parameter of BEC  by a localized dissipative perturbation
(i.e. by  a dissipative defect) was shown in Ref.~\cite{BKPO}.  It was also shown
that the quantum coherence of a strongly interacting BEC loaded in the double-well
trap subject to the phase noise and particle loss can be  completely restored  by
engineering the parameters of the system and controlling the dissipation rate
\cite{WTW1}. Similar results were obtained  with the optical lattices, where  the
particle loss at the boundary acting together with the nonlinearity resulted in
restoration of the coherence and formation of the discrete breathers \cite{WTW2}.
Whereas in the previous  setups the   action of a \textit{localized or boundary}
dissipation  was considered, in the present proposal we consider the action of a
\textit{periodic dissipation} which profoundly affects the   physics in the optical
lattice. Therefore, to distinguish  such a setup from the usual optical lattice, it
will be called below the dissipative lattice. Indeed, a strong periodic dissipation
effectively changes the lattice space group, it results in   a bigger unit cell for
the dissipative lattice as compared  to the same lattice without the periodic
dissipation. The dissipative lattice  can have  both  the  dissipative as well as
the (almost) non-dissipative coherent modes, which are Fourier-like expansions over
some superpositions of the  local modes in each unit cell. The long-term state of
BEC in the dissipative lattice is a lowest energy subset of the effective dark
states, i.e. the states almost unaffected by the dissipation as the result of the
quantum Zeno effect \cite{QZ,CPZeno} (see also the reviews \cite{ZRew1,ZRew2}).
Finally, in contrast to the previous works,  a weak nonlinearity in our case plays
only an auxiliary role (see the honeycomb lattice example below), which is to break
the energy  degeneracy of the effective dark states and   select one particular
state.

The paper is arranged as follows. In section \ref{sec2} we consider the simple case
of the one-dimensional periodic dissipative lattice, where the   action of a
periodically distributed dissipative  sites  is discussed. Then, in section
\ref{sec3}, we consider a more involved case of a dissipative honeycomb optical
lattice. Section \ref{sec4} contains the discussion  of the main results and a
general perspective  on the  dissipative periodic lattices.

\section{The one-dimensional  lattice with a  periodic sublattice of dissipative
sites}
\label{sec2}

\begin{figure}[htb]
\begin{center}
\epsfig{file=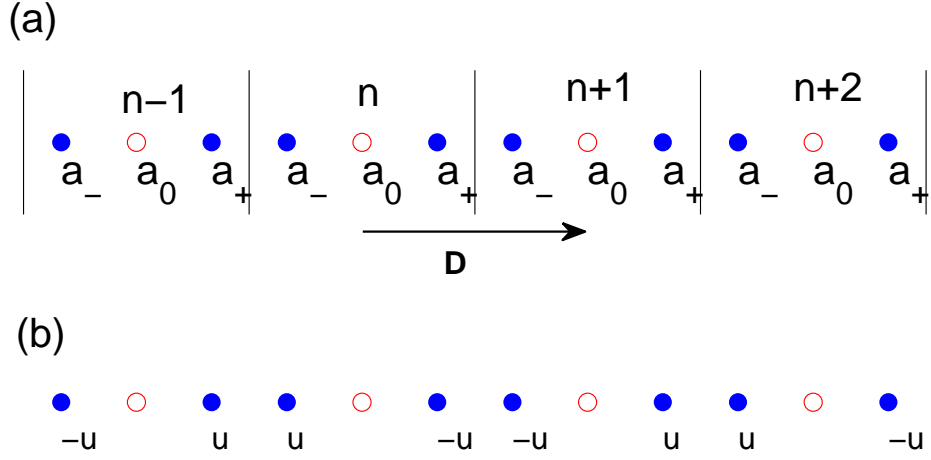,width=0.85\textwidth} \caption{(Color online) (a)  A
schematic depiction of the one-dimensional lattice with a periodic sublattice of
dissipative sites (open circles). The lattice unit cell selection is shown  by
vertical lines, where in each unit cell the lattice sites are enumerated by
``$-$'', ``$0$'' and ``$+$'' (shown by the subscript to the local boson operator
$a$). The consequential   numbers on   top enumerate the unit cells and the vector
labelled by $\bD$ at the bottom  is the dissipative lattice period. (b) Schematic
depiction of the Bloch mode $\varphi_{\frac{L}{2},-}$. Here $u =
\frac{1}{\sqrt{2L}}$. }
\label{FG1}
\end{center}
\end{figure}

Here we consider the simplest case of one-dimensional dissipative lattice to
illustrate the main idea. Specifically, we concentrate on the one-dimensional
lattice where  each third lattice site is dissipative with the same dissipative
rate, see Fig. \ref{FG1}. The selected    periodicity of the dissipative sites is
the most dense one within the class of the dissipative sublattices which still
allow for the non-dissipative coherent modes in the unit cell (see
Fig.~\ref{FG1}(a)). We consider the tight-binding approximation with the nearest
neighbor tunneling and a weak nonlinearity as compared to the tunneling and
dissipation rates.  Then the standard boson Hubbard Hamiltonian applies, which in
the notations of Fig.~\ref{FG1}(a) reads
\begin{eqnarray}
&&H =-J\sum_{n}\{a^\dag_{n,0}(a_{n,-} + a_{n,+}) + a^\dag_{n,+}a_{n+1,-}
+\textrm{h.c.} \} \nonumber\\
&& + \frac{U}{2}\sum_{n}\{(a^\dag_{n,-})^2a^2_{n,-} + (a^\dag_{n,0})^2a^2_{n,0} +
(a^\dag_{n,+})^2a^2_{n,+}\},
\label{EQ1}
\end{eqnarray}
where the  index $n$ enumerates the unit cells of the dissipative lattice. Note
that the dissipative lattice period   is $\bD = 3\bd$, where $\bd$ is the period of
the original (non-dissipative) lattice.

We assume that the atoms are removed with a constant and uniform rate $\Gamma$ from
the sublattice of sites as indicated in Fig. \ref{FG1}, which can be realized, for
example, by the technique of Refs.~\cite{ContRem1,ContRem2}. Then,  the state of
the system is given by the density matrix $\rho$ satisfying the quantum master
equation in the Lindblad form  (see, for more details, Ref.~\cite{NlZeno} and
Ref.~\cite{Book} for a general discussion)
\be
\frac{d \rho}{dt} = -\frac{i}{\hbar}[H,\rho] +
\Gamma\sum_{n}\mathcal{D}\left[a_{n,0}\right]\rho,
\en{EQ2}
where the Lindblad term is defined as $\mathcal{D}[a]\rho \equiv a\rho a^\dag
-\frac12[a^\dag a\rho+\rho a^\dag a]$.

The reason to use a larger unit cell  of the lattice in  Eq. (\ref{EQ1}) and
(\ref{EQ2}) is that,  for a strong dissipation rate, the population of the
dissipative sites is locked to that of the non-dissipative ones (see Ref.
\cite{ThreeSite} for further details) and can be adiabatically eliminated. In this
way, the problem of finding the ground state of BEC in the dissipative lattice is
reduced to that of a modified non-dissipative lattice, where the boson   operators
in each unit cell are some coherent modes over the non-dissipative sites
($c_{n,\pm}$ below). Our use of the introduced notion of the dissipative periodic
lattice leads to a significant reduction of complexity of the analytical analysis.

We  focus on the limit of strong dissipation $\Gamma\gg J/\hbar$ and neglect the
nonlinearity   assuming it  to be too weak to have contribution on the considered
time scale. Specifically, we assume  the condition  $J/(\hbar\Gamma)\gg
U\langle{a^\dag a}\rangle/J\,$ where $\langle a^\dag a\rangle$ is the average
atomic filling of the non-dissipative wells. By performing the adiabatic
elimination of the dissipative sites ($a_{n,0}$) (for the details, consult
Ref.~\cite{QSG}, section II) the master equation is reduced to that for the
non-dissipative sites only (with the reduced density matrix $\rho_R$), which is in
the same form as Eq. (\ref{EQ2}) with, however, a reduced dissipation rate $\gamma$
and a reduced Hamiltonian $H_R$ (the terms depending on the operators of the
dissipative sites are thrown away). The coupling to the dissipative sites in
Eq.~(\ref{EQ1}) suggests to define the coherent modes in each unit cell as follows
$c_{n,\pm} = (a_{n,+}\pm a_{n,-})/\sqrt{2}$, then the tunneling part of the reduced
Hamiltonian becomes
\begin{eqnarray}
H_R &=& \frac{J}{4}\sum_n\{c^\dag_{n,-}c_{n+1,-} - c^\dag_{n,+}c_{n+1,+} +
\mathrm{h.c.} \}\nonumber\\
&+&\frac{J}{2}\sum_n\{c^\dag_{n,+}(c_{n+1,-}-c_{n-1,-}) +\mathrm{h.c.}\}.
\label{EQ3}
\end{eqnarray}
Since only mode $c_{n,+}$ is coupled directly (with the coupling coefficient
$\sqrt{2}J$) to the dissipative mode $a_{n,0}$ in each unit cell,  the reduced
master equation reads (Ref.~\cite{QSG}, section II)
\be
\frac{d \rho_R}{dt} = -\frac{i}{\hbar}[H_R,\rho_R] +
\gamma\sum_{n}\mathcal{D}\left[c_{n,+}\right]\rho_R, \quad \gamma \equiv
\frac{8J^2}{\hbar^2\Gamma}.
\en{EQ4}
For the times  staring from $t\sim 1/\gamma$   the state of the system is defined
by the global coherent  modes which are decoupled from the dissipative modes.
Indeed, let us define the coherent lattice modes (which can be called the
dissipative Bloch modes) as follows
\be
\varphi_{k,\pm} = \frac{1}{\sqrt{L}}\sum_n e^{\frac{2\pi i}{L}kn }c_{n,\pm},
\en{EQ5}
where  $L$ is the total number of   unit cells (i.e. $L = N/3$ where $N$ is the
number of lattice sites, see Fig. \ref{FG1}). The    unitary transformation
(\ref{EQ5}) leaves the  Lindblad term in the master equation (\ref{EQ4}) invariant,
i.e. Eq. (\ref{EQ4}) has the same form in the Bloch basis
\be
\frac{d \rho_R}{dt} = -\frac{i}{\hbar}[H_R,\rho_R] +
\gamma\sum_{n}\mathcal{D}\left[\varphi_{n,+}\right]\rho_R.
\en{EQ6}
Eq. (\ref{EQ6}) defines  a subspace of the dark states, which consists of the Bloch
modes decoupled from the dissipative modes $\varphi_{k,+}$. To find such modes we
rewrite the Hamiltonian (\ref{EQ3}) in   the Bloch basis
\begin{eqnarray}
&& H_R = \frac{J}{2}\sum_k \cos\left(\frac{2\pi}{L}k\right)(\varphi^\dag_{k,-}\varphi_{k,-}
- \varphi^\dag_{k,+}\varphi_{k,+})\nonumber\\
&& +J\sum_k \left\{i\sin\left(\frac{2\pi}{L}k \right)\varphi^\dag_{k,-}\varphi_{k,+} +
\mathrm{h.c.}\right\}.
\label{EQ7}
\end{eqnarray}
It is seen that the   dark subspace contains two  Bloch modes $ \varphi_{0,-}$ and
$\varphi_{\frac{L}{2},-}$ (more precisely, the Bloch modes with $\delta \equiv
(k-k_d)/L\ll 1\,$, for   $k_d =0$ or $ k=L/2$, have significantly reduced  decay
rate due to weak coupling  to the dissipative modes, since the coupling coefficient
reads $J|\sin\left(\frac{2\pi}{L}k \right)|\approx J|\delta| $). Moreover, since
the dissipative lattice modes $\varphi_{k,-}$ have  a negative  kinetic energy, see
Eq. (\ref{EQ7}),   the lowest energy mode is the Bloch mode
$\varphi_{\frac{L}{2},-}$. Note that the long-term ground state
$\varphi_{\frac{L}{2},-}$ breaks the discrete translational symmetry of the
dissipative lattice (the translation by $\bD$), since in terms of the local basis
we have
\be
\varphi_{\frac{L}{2},-} = \frac{1}{\sqrt{2L}}\sum_n (-1)^n(a_{n,+}- a_{n,-}),
\en{EQ8}
which  is schematically depicted  in Fig.~\ref{FG1}(b).

Let us summarize the main conclusions of this simple example. The dissipative
optical lattice with a strong dissipation rate    is effectively equivalent to an
effective  non-dissipative  lattice, which has   unusual properties of the
long-term ground state   of BEC. In this particular case the effectively negative
kinetic energy of the non-dissipative sublattice results in the Bloch state with a
non-zero quasimomentum, which  breaks the translational symmetry of the dissipative
lattice. Below we consider a more intriguing  example of two-dimensional lattice,
where the sublattice of the dissipative sites induces a    long-term state of BEC
with a vortex-like phase distribution.


\section{The honeycomb  optical lattice with a periodic sublattice of  dissipative
sites}
\label{sec3}

\begin{figure}[htb]
\begin{center}
\epsfig{file=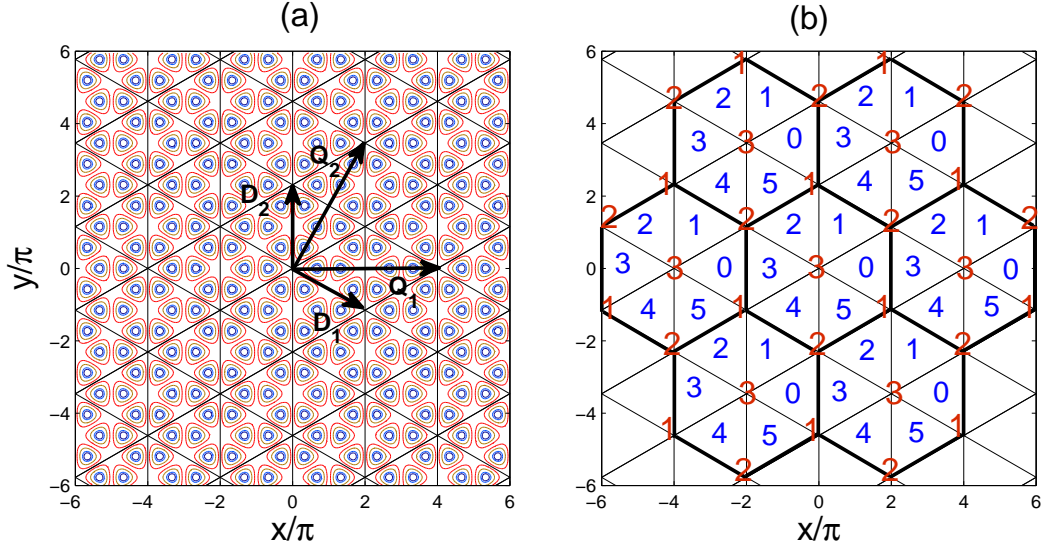,width=0.85\textwidth} \caption{(Color online) (a) Contour
plot of  the honeycomb optical lattice (\ref{E1}) (the lattice minima are shown).
The dissipative sites are located at the centers of the four-site elementary
triangles (shown here  by the division lines). The vectors $\bD_{1,2}$ give the
basis of the dissipative lattice periods, whereas $\bQ_{1,2}$   connect  the
equivalent hexagons (shown by the thick lines in panel (b)). (b) Division of the
dissipative lattice into the set of the  equivalent hexagons used for
representation of the Hamiltonian. The numbers at the vertices  give the indices of
the local boson operators corresponding to the non-dissipative wells,  while the
inner numbers enumerate the triangles inside each equivalent hexagon.}
\label{FG2}
\end{center}
\end{figure}

Let us now consider   the lattice with  the hexagonal symmetry, namely, the
honeycomb optical lattice. The specific choice of the lattice has two reasons.
First, in the tight-binding limit, the honeycomb lattice reduces to just four
nearest-neighbor sites with equal tunneling amplitude, where three sites are
arranged in the form of an equilateral triangle and  one more  cite is placed at
its center, see Fig.~\ref{FG2}(a). From this perspective, by application of the
external dissipation to the central cite in each such triangle one arranges for the
three non-dissipated sites coupled to it, thus the next possible number as compared
to the one-dimensional case considered in section \ref{sec2}. Second, within  the
class of two-dimensional lattices, the lattices with the hexagonal symmetry are
known for their intriguing properties, such as the appearance of the Dirac cones
(i.e. the Dirac Hamiltonian) in the Bloch band intersections and the related
topological phase transitions \cite{Latt2,IntDirCone}, including  the phase
transition from the conventional to unconventional superfluidity \cite{SpinBos}.
Thus, one naturally expects the dissipative honeycomb lattice to possess an
interesting long-term ground state.

The simplest honeycomb optical lattice, which was experimentally realized
\cite{Latt1}, is created by intersection of three laser beams of of the same
frequency and equal intensities. When the lasers are blue detuned from the atomic
transition frequency, the lattice minima are at the vertices of honeycombs. In this
case, the optical lattice potential reads $V(\bx) = V_0\sum_{j=1}^3 \cos(\bb_j
\bx)$, where $\bx = (x,y)$, $\bb_{1,2}$ are the  basis vectors of the reciprocal
lattice, with  $\bb_2$ being the vector $\bb_1$ rotated by $\pi/3$,   and $\bb_3 =
\bb_2 - \bb_1$. We  note that, interestingly, the following  lattice
\be
V = 8V_0\sin^2\left(\frac{\bb_1\bx}{2}\right)
\sin^2\left(\frac{\bb_2\bx}{2}\right)\sin^2 \left(\frac{\bb_3\bx}{2}\right),
\en{E1}
shown in  Fig.~\ref{FG2}(a),  has also the honeycomb shape, but possesses much more
pronounced minima of the lattice wells (it can be realized with the recent optical
lattice technology \cite{Latt3}, see also Ref. \cite{OptExpr}). The results
obtained below, however,   apply to any of the optical lattices reducible in the
tight-binding limit to the honeycomb   arrangement of the lattice wells,   where in
each equilateral triangle there are three non-dissipated sites    located at the
vertices and the dissipative site   located  at the center, i.e. as in
Fig.~\ref{FG2}.

Taking into account the discussion of section \ref{sec2},  we will use from the
start the nomenclature defined by the dissipative lattice, see Fig. \ref{FG2}(b).
In particular, the lattice wells are enumerated  as shown in Fig.~\ref{FG2}, where
each  elementary triangle (shown by the thin lines in Fig. \ref{FG2}(a)) contains
one dissipative lattice well at the center, to which we assign the boson
annihilation operator $a_0$, and three non-dissipative wells at the vertices, with
the corresponding operators $a_{1,2,3}$, as is schematically shown  in
Fig.~\ref{FG2}(b). It is seen that the dissipative lattice periods are given as
$\bD_1=\frac{8\pi}{3|\bb|^2}(2\bb_1-\bb_2)$ and
$\bD_2=\frac{8\pi}{3|\bb|^2}(2\bb_2-\bb_1)$ (which are twice the original honeycomb
lattice periods). It proves convenient to use the equivalent hexagons shown by the
thick lines in Fig.~\ref{FG2}(b)   as the unit cells of the dissipative lattice
(see below). Therefore, we introduce the basis of the translations $\bQ_1 = 2\bD_1
+\bD_2$ and $\bQ_2 = 2\bD_2 +\bD_1$, Fig.~\ref{FG2}(a), which translate the
equivalent hexagons into each other without any change in the enumeration of the
lattice wells. The annihilation operator at a given lattice site now is denoted by
$a_{\bq,n,l}$, where $l=0,1,2,3$ is the lattice site index inside the triangle,
$n\in\{0,\ldots,5\}$ is the triangle index and  $\bq = q_1\bQ_1 +q_2 \bQ_2$, $q_j
\in \{0,\pm1,\pm2, \ldots\}$, is the hexagon index.

Our  division of the   hexagonal lattice into the equivalent hexagons, as in
Fig.~\ref{FG2}(b),  and   the specific  enumeration scheme    are dictated by the
geometry of the strongly dissipative sublattice. Though the  indices  of the local
operators assigned to the non-dissipative sites in each hexagon still can be
distributed at will, the selected enumeration  shown in Fig.~\ref{FG2}(b) proves to
be the most convenient.

 In the tight-binding
limit with the nearest-neighbor tunneling confined to the lowest Bloch band, the
boson Hubbard Hamiltonian   reads
\be
H =-J\sum_{\langle{,}\rangle}(a^\dag_{\bq,n,k}a^{}_{\bq^\prime,m,l}+\textrm{h.c.})
+ \frac{U}{2}\sum_{\bq,n,k}(a^\dag_{\bq,n,k})^2a^2_{\bq,n,k},
\en{E2}
where the first sum is over the nearest neighbors and $U$ is the nonlinear
interaction strength proportional to the $s$-wave scattering length of BEC.

We assume that the standard Markovian dissipation with the constant and spatially
uniform rate $\Gamma$ is applied to the lattice wells located at the centers of the
elementary triangles of Fig.~\ref{FG2}. Then the master equation reads
\be
\frac{d \rho}{dt} = -\frac{i}{\hbar}[H,\rho] +
\Gamma\sum_{\bq}\sum_{n=0}^5\mathcal{D}\left[a_{\bq,n,0}\right]\rho.
\en{E3}
In the limit of strong dissipation, i.e.
\be
\Gamma\gg \frac{J}{\hbar},\qquad \Gamma\gg \frac{U\langle{a^\dag a}\rangle}{\hbar},
\en{Cond1}
the dissipative wells can be  adiabatically eliminated, thus reducing  the master
equation (\ref{E3}) to that for the non-dissipative wells only. Skipping the
details (see Ref.~\cite{QSG}, section II), let us write down the resulting reduced
master equation for the density matrix $\rho_R$ describing  the non-dissipative
sites
\be
\frac{d \rho_R}{dt} = -\frac{i}{\hbar}[H_R,\rho_R] + \mathcal{L}\rho_R, \quad
\mathcal{L}\rho_R=\gamma\sum_{\bq}\sum_{n=0}^5\mathcal{D}[c_{\bq,n,3}]\rho_R,
\en{E4}
where the reduced rate reads $\gamma = {12J^2}/({\hbar^2\Gamma})$, $H\equiv H_R$ is
the reduced Hamiltonian (the part of $H$ from  Eq.~(\ref{E1})  dependent only on
$a_{\bq,n,l}$ and $a^\dag_{\bq,n,l}$ with  $l=1,2,3$) and  the local coherent basis
used in the representation of Eq.~(\ref{E4})  is defined by the unitary
transformation
\be
c_{\bq,n,l} = \frac{1}{\sqrt{3}}\sum_{k=1}^3e^{i\frac{2\pi}{3}lk}a_{\bq,n,k}, \quad
l=1,2,3.
\en{E5}
The reduced Hamiltonian $H_R$   is obtained from the Hamiltonian (\ref{E2}) by
throwing away the terms with operators ($a_{\bq,n,0}$ and $a^\dag_{\bq,n,0}$)
belonging to the dissipative wells. Substituting from Eq.~(\ref{E5}) we obtain it
in the form
\begin{widetext}
\begin{eqnarray}
H_R &=& -\frac{2J}{3}
\sum_{\bq}\biggl[\sum_{s=0}^2\left\{\vec{c}^\dag_{\bq,2s}T\vec{c}_{\bq,2s+1} +
\vec{c}^\dag_{\bq,2s+1}T^*\vec{c}_{\bq,2s+2} + \textrm{h.c.}\right\}  \nonumber\\
&& + \sum_{n=0}^5
\left\{ \vec{c}^\dag_{\bq,n}\Lambda\vec{c}_{\bq+\br_n,n+3}+
\textrm{h.c.}\right\}\biggr] + \sum_{\bq}\sum_{n=0}^5 H^{(int)}_{\bq,n},
\label{E6}\end{eqnarray}
\end{widetext}
where  the vector notation  \mbox{$\vec{c}^\dag_{\bq,n} =
(c^\dag_{\bq,n,1},c^\dag_{\bq,n,2},c^\dag_{\bq,n,3})$} is used. The coupling
matrices are given as follows \mbox{$T_{l,k} =
\cos(\frac{\pi}{3}[l-k])e^{-i\pi(l-k)/3}$,} $T^*$ is its complex conjugate and
$\Lambda_{l,k} = \frac12\cos(\frac{2\pi}{3}[l-k])$. The shift vector $\br_n$ in the
second  sum  of Eq. (\ref{E6}) is defined as follows (see Fig.~\ref{FG2})
\[
\br_0 = \bQ_1, \; \br_1 = \bQ_2,\; \br_2 = \bQ_2-\bQ_1,\; \br_{n+3} = - \br_n.
\]
The nonlinear interaction term in Eq. (\ref{E6}) is given in  terms of the local
coherent basis $c_{\bq,n,l}$, where the  term $H^{(int)}_{\bq,n}$ describes the
nonlinear interaction in  the elementary triangle $(\bq,n)$. It has  the following
form (omitting the index $(\bq,n)$ in $c_{\bq,n,l}$ for simplicity)
\begin{eqnarray}
H^{(int)}_{\bq,n} &=& \frac{U}{6}\biggl\{\sum_{k=1}^3 (c^\dag_k)^2c_k^2 +
4\sum_{k<l}c^\dag_k c^{}_k c^\dag_l c^{}_l \nonumber\\
&&
+ 2(c^\dag_1c^\dag_2c_3^2 + c^\dag_2c^\dag_3c_1^2 +c^\dag_1c^\dag_3c_2^2 +
\textrm{h.c.}) \biggr\}.
\label{E7}\end{eqnarray}

We assume that the nonlinear interaction time scale is much larger than that of the
reduced dissipation, i.e. $t_{int}\sim \hbar/(U\langle c^\dag c\rangle) \gg
t_{diss}\sim 1/\gamma$, which implies  the condition
\be
\frac{U\langle a^\dag a\rangle}{J} \ll \frac{J}{\hbar \Gamma}
\en{Cond2}
(note that the second condition in Eq.~(\ref{Cond1}) follows from condition
(\ref{Cond2}) and  the first condition in Eq.~(\ref{Cond1})). Then the reduced
Lindblad operator $\mathcal{L}$ of Eq.~(\ref{E4}) together with the  reduced
Hamiltonian (\ref{E6}) define the long-time state $|\Psi\rangle$ of the condensate.
First of all, $|\Psi\rangle$  must be the dark state of $\mathcal{L}$:
$\mathcal{L}|\Psi\rangle = 0$. Indeed, this long-time state $|\Psi\rangle$ will be
still subject to the weak dissipation coming from  the nonlinear interaction in Eq.
(\ref{E7}) between the dissipative coherent modes $c_{\bq,n,3}$ and the coherent
modes $c_{\bq,n,1}$ and $c_{\bq,n,2}$ inside each elementary triangle. The form of
the nonlinear dissipation can be found by a similar adiabatic elimination procedure
as above. However, since the rate of the reduced  dissipation is proportional to
the square of the interaction term (see Ref.~\cite{QSG}), the rate of the nonlinear
dissipation is proportional to the square of the  interaction parameter, i.e.
$\gamma_{int} \sim \frac{(U\langle c^\dag c\rangle)^2}{\hbar^2\gamma}$. But then,
under   condition (\ref{Cond2}), the nonlinear dissipation can be neglected since
its time scale is much larger even than the interaction time, we have
$1/\gamma_{int} \sim t^2_{int}/t_{diss}\gg t_{int}$, where $t_{diss}$ is the
reduced linear dissipation time scale   and $t_{int}$ is the interaction time.

Since the dissipative terms due to the nonlinear coupling can be  neglected, the
long-time state $|\Psi\rangle$ is a member of the dark states of $\mathcal{L}$. The
dark subspace is characterized by  the set of   lattice operators \textit{linearly}
decoupled from the dissipative operators $c_{\bq,n,3}$ entering the Lindblad
operator $\mathcal{L}$ (\ref{E4}). To find these operators  let us first
diagonalize as much as possible of the quadratic part of Hamiltonian $H_R$
(\ref{E6}) while keeping the Lindblad operator (\ref{E4}) diagonal. This  amounts
to introducing the dissipative coherent lattice modes, i.e. the dissipative Bloch
waves, by the following unitary transformation (note that there are exactly three
types of different dissipative Bloch  modes, which  we combine in the vector
notation)
\be
\vec{\varphi}_{\bp,k} = \frac{1}{\sqrt{6}\mathcal{N}}\sum_\bq\sum_{n=0}^5
\exp\biggl\{i[\bp\bq +\frac{\pi}{3}kn]\biggr\}\vec{c}_{\bq,n}.
\en{E8}
Here $\mathcal{N}$ is the number of the equivalent  hexagons in each of the two
directions $\bQ_{1,2}$ (see Fig.~\ref{FG2}) and   $\bp$ is the Bloch index, i.e.
$\bp = \frac{1}{{\mathcal{N}}}(p_1\bP_1 + p_2\bP_2)$ with  \mbox{$p_{1,2}
\in\{0,1,2,\ldots,\mathcal{N}-1\}$}, while  the reciprocal lattice periods
$\bP_{1,2}$ are defined by $\bP_i\bQ_j = {2\pi}\delta_{i,j}$. They read  $\bP_1 =
\frac{1}{3}(\bb_1 - \frac12\bb_2)$ and $\bP_2 = \frac{1}{3}(\bb_2 - \frac12\bb_1)$.

The   unitary transformation  given by  Eq.~(\ref{E8})    keeps the  diagonal  form
of the  Lindblad operator $\mathcal{L}$  in Eq.~(\ref{E4})   invariant, i.e. we
have in the Bloch basis
\be
\mathcal{L}\rho_R = \gamma\sum_\bp\sum_{k=0}^5
\mathcal{D}[\varphi_{\bp,k,3}]\rho_R,
\en{E9}
where  the sum is now over the  Bloch  modes.  Substituting Eq. (\ref{E8}) into the
quadratic (i.e. tunneling) part of the Hamiltonian (\ref{E6}) we obtain it in the
form
\begin{widetext}
\be
H^{(tunl)}_R =-\frac{4J}{3}\sum_\bp\left\{\sum_{k}^5\cos\left(\frac{\pi k
}{3}\right)\vec{\varphi}^\dag_{\bp,k}\frac{T+T^*}{2}\vec{\varphi}_{\bp,k} +
\sum_{k_1=0}^5\sum_{k_2=0}^5S_{k_1,k_2}(\bp)\vec{\varphi}^\dag_{\bp,k_1}\Lambda\vec{\varphi}_{\bp,k_2}
\right\},
\en{E10}
\end{widetext}
where $S_{k,l}$ is defined as
\be
S_{l,m}(\bp) =e^{\frac{i\pi}{2}(l+m)} \frac{1}{3}\sum_{n=0}^2
e^{\frac{i\pi}{3}(l-m)n}\cos\left(\frac{\pi}{2}[l-m]+\bp\br_n\right).
\en{E11}
One can easily check the invariance of the Hamiltonian $H_R$ (\ref{E6}) and the
expression for its quadratic part $H^{(tunl)}_R$ (\ref{E10}) with respect to the
dissipative lattice translation group, keeping in mind that a shift by $\bD_{1,2}$
should be supplemented by a change in the operator  indices. The latter amounts to
the unitary transformation: $c_{\bq,n,k} \to e^{-ik\theta}c_{\bq,n,k}$, where
$\theta = \frac{2\pi}{3}$, i.e. to the permutation of the enumeration of the wells
inside each triangle according to the cycle $(1,2,3)\to (2,3,1)$, see
Fig.~\ref{FG2}(b).

Using  the  representation given by Eq.~(\ref{E10}) one can find the Bloch indices
$(\bp,k)$ of the non-dissipated waves. Indeed, the operators $\varphi_{\bp,k,1}$
and $\varphi_{\bp,k,2}$ which decouple from the dissipative operators
$\varphi_{\bp,k,3}$, are solutions of the   following equation (for $l=1,2$)
\be
\cos\left(\frac{\pi k_1 }{3}\right)\frac{T_{l,3}+T^*_{l,3}}{2}\delta_{k_1,k_2} +
S_{k_1,k_2}(\bp)\Lambda_{l,3} = 0.
\en{E12}
Since the first term in Eq.~(\ref{E12}) is diagonal in $(k_1,k_2)$, the matrix
$S_{k_1,k_2}(\bp)$ must be diagonal too. Expression (\ref{E11}) gives only one case
$\bp=0$, i.e. the quasi-momentum must be zero. Then Eq.~(\ref{E12}) reduces to a
simple scalar equation with just two solutions for the second Bloch index: $k=0$ or
$k=3$. Therefore, the four Bloch waves which belong to the dark subspace  of the
linear dissipation  correspond to the coherent operators $\varphi_{\ze,0,l}$ and
$\varphi_{\ze,3,l}$ with $l=1,2$ in Eq. (\ref{E8}). They  have the Bloch energies
$E(\bp=\ze,k=0,l) = -2J$ and $E(\bp=\ze,k=3,l) = 2J$, respectively. Therefore the
lowest energy is doubly degenerate. We note that such a  dark subspace is an
attribute of the dissipative periodic structure and would be impossible, for
instance, with just one equivalent hexagon, since in this case the second term in
Eq. (\ref{E12}) is absent.

\begin{figure}[htb]
\vskip 1cm
\begin{center}
\epsfig{file=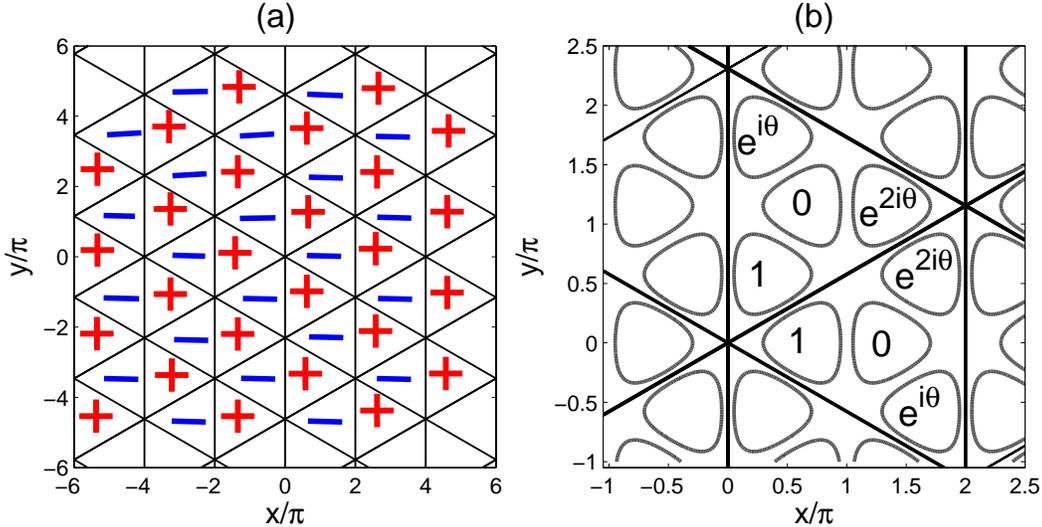,width=0.85\textwidth} \caption{(Color online) Schematic
depiction of the discrete  vortex structure of the superfluid state
$|\Psi_+\rangle$ (\ref{E14}). (a) The distribution of the vortices and
anti-vortices in the elementary triangles  of Fig.~\ref{FG2}. (b) The distribution
of the discrete vortex amplitude and phase between the wells (shown by contours) of
the elementary triangles, with just two being shown, one containing the vortex and
the other one -- the anti-vortex. Here $\theta=\frac{2\pi}{3}$. }
\label{F2}
\end{center}
\end{figure}

Due to the double degeneracy, the actual state of the system  is defined, of
course, by the nonlinear interaction part of the Hamiltonian $H_R$ (\ref{E6}). In
our approximation, given by Eqs.~(\ref{Cond1}) and (\ref{Cond2}),  we have
$U\langle a^\dag a\rangle/J \ll 1$. Thus one can  keep only the degenerate modes in
the nonlinear term, which we denote as $\varphi_+ \equiv \varphi_{\ze,0,1}$ and
$\varphi_-\equiv \varphi_{\ze,0,2}$. Projecting on the degenerate subspace, we
obtain
\be
H^{(int)} =\frac{U}{(6\mathcal{N})^2}\left\{(\varphi^\dag_+)^2\varphi^2_+ +
(\varphi^\dag_-)^2\varphi^2_-
+\varphi^\dag_+\varphi_+\varphi^\dag_-\varphi_-\right\}.
\en{E13}
A simple analysis  of Eq. (\ref{E13}) shows that the interaction energy is
minimized when all atoms are in either one of the two Bloch modes $\varphi_\pm$.
Each of the corresponding Bloch waves is a coherent superposition  of the
alternating discrete vortices and anti-vortices,   each occupying just one of the
elementary triangles, with the vortex centers being at the dissipative sites, as is
easily seen from Eqs. (\ref{E5}) and (\ref{E8}), see Fig.~\ref{F2}. In the explicit
form
\be
|\Psi_\pm\rangle\propto\sum_\bq\sum_{n=0}^5\{e^{\pm i\theta}|G_{\bq,n,1}\rangle
+e^{\mp i\theta}|G_{\bq,n,2}\rangle + |G_{\bq,n,3}\rangle \},
\en{E14}
where $\theta = 2\pi/3$ and $|G_{\bq,n,l}\rangle$ is the ground state in the
respective lattice well.

We note that though a macroscopic  superposition of the two degenerate Bloch waves
$|\Psi_\pm\rangle$ is permitted by the model, one must recall that the discarded
terms of the full Hamiltonian and the nonlinear dissipation will eventually destroy
the macroscopic coherence, thus  one of the two superfluid states will be
spontaneously selected. For instance,  only the coherent superposition $\propto
|\Psi_+\rangle - |\Psi_-\rangle$ with only three bosons in the system is the only
null state of the nonlinear dissipation. Thus, the long-time state of BEC in the
dissipative honeycomb optical lattice of Fig.~\ref{FG2} has the complex-valued
order parameter with the vortex-like phase distribution.


\section{Conclusion}
\label{sec4}

In conclusion, we have introduced  the notion of the dissipative optical lattice as
the optical lattices with a sublattice of dissipative sites and have demonstrated
its utility for engineering of the superfluid states of BEC with the complex-valued
order parameter. The tight-binding approximation and the limit of a strong
dissipation was used to allow for an analytical approach, since  the numerical
simulations would be difficult to carry out for the many-body open system on a
lattice. We have found that the strong dissipation effectively changes the lattice
space group, which results in a larger unit cell   than that of the original
optical lattice without the dissipation.   Moreover, the base boson operators in
the case of the dissipative lattice are linear superpositions of the local
operators of the non-dissipative sites, what results in the  ground state with a
non-trivial complex-valued  order parameter. For instance, in the dissipative
honeycomb lattice the long-time ground state is a coherent superposition of the
alternating discrete vortices and anti-vortices.

We have considered just two examples, the simple one-dimensional case and the
honeycomb lattice,  where we have chosen some particular distributions of the
dissipative sites which allow one to obtain the long-term ground state in  explicit
analytical form. However, similar results are expected to hold quite generally for
the optical lattices with a sublattice of dissipative sites.  Moreover,  similar
behavior of BEC is expected in   dissipative optical lattices in the continuous
limit (as opposed to the tight-biding limit), which will be considered in a future
publication.

\acknowledgments   This work was supported by  the FAPESP and CNPq of Brazil.

\end{document}